\documentclass[10pt,conference]{IEEEtran}
\usepackage{epsfig,rotating,setspace,latexsym,amsmath,epsf,amssymb,amsfonts,bm,theorem,cite,enumerate,longtable,accents}
\usepackage{algorithm,algorithmic,graphicx,epsf,authblk,epstopdf,url,color,multirow,longtable}
\usepackage{mathtools}
\DeclarePairedDelimiter{\ceil}{\lceil}{\rceil}

\newtheorem{theorem}{Theorem}

\newtheorem{corollary}{Corollary}

\newtheorem{remark}{Remark}

\newtheorem{example}{Example}
\newenvironment{Proof}[1]{\medskip\par\noindent{\bf Proof:\,}\,#1}{{\mbox{\,$\blacksquare$}\par}}

\IEEEoverridecommandlockouts
\allowdisplaybreaks

\title{Communication Cost of Two-Database Symmetric Private Information Retrieval: A Conditional Disclosure of Multiple Secrets Perspective\thanks{This work was supported by ARO Grant W911NF2010142, and NSF Grants CCF 17-13977 and ECCS 18-07348.}}

\author{Zhusheng Wang \qquad Sennur Ulukus\\
	\normalsize Department of Electrical and Computer Engineering\\
	\normalsize University of Maryland, College Park, MD 20742\\
	\normalsize  \emph{zhusheng@umd.edu} \qquad \emph{ulukus@umd.edu}}

\begin{document}
\date{}
\maketitle

\begin{abstract}
We consider the total (upload plus download) communication cost of two-database symmetric private information retrieval (SPIR) through its relationship to conditional disclosure of secrets (CDS). In SPIR, a user wishes to retrieve a message out of $K$ messages from $N$ non-colluding and replicated databases without learning anything beyond the retrieved message, while no individual database learns the retrieved message index. In CDS, two parties each holding an individual input and sharing a common secret wish to disclose this secret to an external party in an efficient manner if and only if their inputs satisfy a public deterministic function. As a natural extension of CDS, we introduce conditional disclosure of multiple secrets (CDMS) where two parties share multiple i.i.d.~common secrets rather than a single common secret as in CDS. We show that a special configuration of CDMS is equivalent to two-database SPIR. Inspired by this equivalence, we design download cost efficient SPIR schemes using bipartite graph representation of CDS and CDMS, and determine the exact minimum total communication cost of $N=2$ database SPIR for $K=3$ messages.
\end{abstract}

\section{Introduction}
As initially introduced in \cite{SPIR_ORI}, symmetric private information retrieval (SPIR) refers to the problem where a user downloads a message out of $K$ possible messages stored in $N$ non-colluding and replicated databases in such a way that, not only no individual database can know which message the user has just downloaded, but also the user learns nothing about the remaining messages stored in the databases. The total communication cost of SPIR consists of two parts: the total number of bits sent from the user to the databases (\emph{upload cost}) denoted by $U$, and the total number of bits downloaded by the user from the databases (\emph{download cost}) denoted by $D$. For a message length of $L$ bits, the total communication cost $(U+D)$ of SPIR depends on three basic parameters $(N,K,L)$.
 
In \cite{SPIR}, without any constraints on $U$ and $L$, the optimal download cost for SPIR is found to be $\frac{NL}{N-1}$, which does not depend on $K$. In \cite{Min_Uploadcost_SPIR}, without any constraints on $D$ and $L$, the optimal upload cost for SPIR is found to be $\log_2 (\ceil{K^\frac{1}{N-1}})$ which does not depend on $L$. In addition, \cite{PSI_journal, MP-PSI_journal, SPIR_atPIR, SPIR_Eavesdropper, SPIR_Mismatched, SPIR_coded, SPIR_Collusion} explore the optimal download cost of SPIR under various extended conditions without a consideration on the upload cost (see also many other important variants of PIR and SPIR in \cite{PIR,Tian_upload,one_extra_bit,MMPIR,PIR_coded,Coded_PIR_Server,Kumar_PIRarbCoded,PIR_lifting,ChaoTian_coded_minsize,Karim_nonreplicated,Tamo_journal,PrivateComputation,AsymmetryHurtsPIR,NoisyPIR,PIR_WTC_II,PrivateSearch,PIR_Collusion,SemanticPIR,SDB_PIR,SDB_MMPIR2,PIR_cache_edge,tandon_cache_2017,Cache-aided_PIR,PrefetchingPIR,PartialPSI_PIR,StorageConstrainedPIR_Wei,PIR_PSI,MMPIR_PrivateSideInfo,ChaoTian_leakage,leakyTandonJournal,securestoragePIR,XSTPIR,XSTPIR_MDS,PIR_decentralized,HeteroPIR,tandon-attia,efficient_storage_ITW2019,StorageCost}). To the best of our knowledge, our paper is the first one to investigate the overall (upload and download) communication cost of SPIR with a particular focus on $L = 1$ in an information-theoretic setting. Our focus on $L=1$ is motivated by two observations: First, as pointed out in \cite{PIR}, when $L$ is allowed to approach infinity, download cost dominates the upload cost, and the consideration of total cost becomes trivial. Second, in some cryptographic applications, e.g., \cite{PSI_journal, MP-PSI_journal}, only $L = 1$ may make practical sense. 

As a classical cryptographic primitive, conditional disclosure of secrets (CDS) is first introduced in \cite{SPIR_ORI} as well to help devise an achievable SPIR scheme. Since CDS itself functions as an essential building block in applications such as secret sharing and attribute based encryption \cite{graph_based_ss, fuzzy_encryption, CDS_property}, CDS has also attracted significant attention as a stand-alone computer science problem. Recently, information-theoretic CDS is formulated in \cite{CDS,linearCDS} to characterize the maximum number of secret bits that can be securely disclosed per communication bit whenever a pre-defined condition is satisfied.

In this paper, we first show the equivalence between a special CDMS configuration and the two-database SPIR. Following this equivalence, we investigate the total communication cost of two-database SPIR through the characteristics of CDS and CDMS. We utilize CDS/CDMS to determine an upload cost, and then proceed to minimize the download cost for the given fixed upload cost. We then consider the feasible upload and download cost achievable region. In the example of $K = 3$ and $L = 1$, we find two optimal corner points for the upload and download cost pair. These two corner points outperform the best-known results in the literature \cite{SPIR, Min_Uploadcost_SPIR} and lead to the optimal total communication cost.

\section{Problem Formulation}
\subsection{Symmetric Private Information Retrieval}
Following the classical SPIR problem statement in \cite{SPIR}, we consider $N \geq 2$ non-colluding databases with each individual database storing the replicated set of $K \geq 2$ i.i.d.~messages $W_{1:K}$. Moreover, $L$ i.i.d.~symbols within each message are uniformly selected from a sufficiently large finite field $\mathbb{F}_q$, 
\begin{align}
    H(W_k) &= L, \quad \forall k \label{Message Length} \\
    H(W_{1:K}) &= H(W_1) + \dots + H(W_K)  = KL \label{Message IID}
\end{align}

A random variable $\mathcal{F}$ is used to denote the randomness of the retrieval strategy selection implemented by the user. Due to the user privacy constraint, the realization of $\mathcal{F}$ is only known to the user, and unknown to any of the databases. Due to the database privacy constraint, databases need to share some amount of common randomness $\mathcal{R}$.

The message set $W_{1:K}$ stored in the databases is independent of the desired message index $k$, retrieval strategy randomness $\mathcal{F}$ and common randomness $\mathcal{R}$,  
\begin{align} \label{Message Set Independence}
    I(W_{1:K};k,\mathcal{F},\mathcal{R}) = 0, \quad \forall k
\end{align}

Using the desired message index, the user generates a query for each database according to the retrieval strategy randomness $\mathcal{F}$. Hence, the queries $Q_n^{[k]}, n \in [N]$ are deterministic functions of $\mathcal{F}$, 
\begin{align} \label{Deterministic Query}
    H(Q_1^{[k]},\dots,Q_N^{[k]}|\mathcal{F}) = 0, \quad \forall k
\end{align}

After receiving a query from the user, each database should respond with a truthful answer based on the stored message set and common randomness, 
\begin{align} \label{Deterministic Answer} 
   \!\!\!\text{[deterministic answer]} \;  H(A_n^{[k]}|Q_n^{[k]},W_{1:K},\mathcal{R}) = 0, \forall n,  \forall k \!\!
\end{align}

After collecting all $N$ answers from the databases, the user should be able to decode the desired messages $W_{k}$ reliably, 
\begin{align} \label{Reliability} 
  \text{[reliability]} \quad &H(W_k|\mathcal{F},A_{1:N}^{[k]}) = 0,  \quad \forall k
\end{align}

Due to the user privacy constraint, the query generated to retrieve the desired message should be statistically indistinguishable from other queries, thus, for all $k^\prime \neq k, k^\prime \in [K]$,
\begin{align} \label{User Privacy}
    \text{[user privacy]} \quad  (Q_n^{[k]},&A_n^{[k]},W_{1:K},\mathcal{R}) \notag \\
    &\sim ~ (Q_n^{[k^\prime]},A_n^{[k^\prime]},W_{1:K},\mathcal{R}) 
\end{align}

Due to the database privacy constraint, the user should learn nothing about  $W_{\bar{k}}$ which is the complement of $W_{k}$, i.e., $W_{\bar{k}} = \{W_1,\cdots,W_{k-1},W_{k+1},\cdots,W_K\}$, 
\begin{align}
    \text{[database privacy]} \quad I(W_{\bar{k}};\mathcal{F},A_{1:N}^{[k]}) = 0,  \quad \forall k \label{Database Privacy} 
\end{align}

An achievable SPIR scheme is a scheme that satisfies the reliability constraint \eqref{Reliability}, the user privacy constraint \eqref{User Privacy} and the database privacy constraint \eqref{Database Privacy}. In this paper, we focus on the overall communication cost, which is a sum of the number of uploaded bits (named \emph{upload cost} and denoted by $U$) and the number of downloaded bits (named \emph{download cost} and denoted by $D$), within the retrieval scheme. As a consequence, the most efficient achievable scheme is the scheme with the lowest total communication cost, i.e., the one that achieves $C^*=\inf (U+D)$ over all achievable SPIR schemes.

\subsection{Conditional Disclosure of a Secret}
Two parties Alice and Bob possess their respective inputs $X, Y$ and share a common secret $S$. Alice and Bob also share an independent randomness $\mathcal{R}$ to assist the secret disclosure of $S$. With the knowledge of the inputs $X,Y$ but without knowing the common randomness $\mathcal{R}$, another party Carol wishes to learn the secret $S$ under a specific condition by communicating with Alice and Bob simultaneously. Generally, this condition is described as a deterministic public function. Specifically, given a globally public function $f$, the secret $S$ is disclosed to Carol if and only if $f(X,Y) = 1$ is true. By contrast, if $f(X,Y)$ is not equal to $1$, no information about the secret $S$ should be revealed to Carol. To that end, Alice sends a signal $A_X$ and Bob sends another signal $B_Y$ to Carol. 

The signals are determined by all the information contained in Alice or Bob before being sent to Carol,
\begin{align}
    \text{[deterministic signal]} \quad H(A_X|X,S,\mathcal{R}) &= 0 \notag \\
    H(B_Y|Y,S,\mathcal{R}) &= 0
\end{align}

If the condition is satisfied, Carol is able to decode the secret by using all the information she possesses,
\begin{align}
    \text{[validity]} \;\; H(S|X,Y,A_X,B_Y) = 0, \;\; \text{if $f(X,Y) = 1$}
\end{align}
Otherwise, if the condition is not satisfied, Carol cannot learn anything about the secret based on all the information she has,
\begin{align}
    \text{[security]} \;\; I(S;X,Y,A_X,B_Y) = 0, \;\; \text{if $f(X,Y) \neq 1$}
\end{align}

The information-theoretic objective of CDS is to minimize the number of bits contained in $A_X$ and $B_Y$.

\subsection{Conditional Disclosure of Multiple Secrets}
Here, we introduce the concept of CDMS as an extension of CDS. Two parties Alice and Bob possess their respective inputs $X, Y$ and share $K$ i.i.d.~common secrets denoted by $S_1, \dots, S_K$. Alice and Bob also share an independent randomness $\mathcal{R}$ to assist the secret disclosure. With the knowledge of the inputs but without knowing the common randomness, another party Carol expects to learn partial secrets under some specific conditions (one for each secret) by communicating with Alice and Bob simultaneously. Specifically, assuming that a sequence of functions $f_k, k \in [K]$ are globally public, then for all $k \in [K]$, the secret $S_k$ is disclosed to Carol if and only if her corresponding condition $f_k(X,Y)$ is equal to $1$. Otherwise, no information about the secret $S_k$ should be revealed to Carol. To that end, Alice and Bob send integrated signals $A_X$ and $B_Y$, respectively, to Carol. As a result, the constraints in CDMS generalize to the following ones.

The integrated signals are determined by all the information contained in Alice or Bob before being sent to Carol,
\begin{align}
    \text{[deterministic signal]} \quad H(A_X|X,S_{1:K},\mathcal{R}) &= 0 \notag \\
    H(B_Y|Y,S_{1:K},\mathcal{R}) &= 0
\end{align}

For all $k \in [K]$, if the condition $f_k$ is satisfied, Carol is able to decode the secret $S_k$,
\begin{align}
    \text{[validity]} \; H(S_k|X,Y,A_X,B_Y) = 0, \; \text{if $f_k(X,Y) = 1$}
\end{align}
For all $k \in [K]$, if the condition $f_k$ is not satisfied, Carol learns nothing about the secret $S_k$,
\begin{align}
    \text{[security]} \; I(S_k;X,Y,A_X,B_Y) = 0, \; \text{if $f_k(X,Y) \neq 1$}
\end{align}

Likewise, the information-theoretic objective of CDMS is to minimize the number of bits contained in $A_X$ and $B_Y$.

\section{Main Results}
We design the particular CDMS configuration given below:
\begin{enumerate}
    \item First, Carol selects a random index $k$, which is uniformly distributed over the set $[K]$; $k$ is independent of the secrets as well as common randomness in Alice and Bob.
    \item Second, Carol selects two random vectors $X$ and $Y$ such that no information about $k$ is leaked in the individual vectors $X$ or $Y$. 
    \item Third, Carol sends $X$ to Alice and $Y$ to Bob. 
    \item Globally known condition functions are set in accordance with the selection of random vectors $X$ and $Y$, such that, at all times only one condition function $f_k$ can be $1$.
\end{enumerate}

\begin{theorem}
CDMS configured as above is equivalent to SPIR with two replicated and non-colluding databases.
\end{theorem}

\begin{Proof}
Within the given configuration, Alice and Bob can be treated as database 1 and database 2, and Carol as the user; the secrets $S_{1:K}$ can be treated as the message set $W_{1:K}$; the random variable $k$ as the desired message index at the user; the inputs $X,Y$ as the queries $Q_1^{[k]},Q_2^{[k]}$; and the signals $A_X,B_Y$ as the answers $A_1^{[k]},A_2^{[k]}$. Thus, we have the following conversions, which complete the proof:
\begin{enumerate}
    \item \emph{Deterministic signal} becomes \emph{deterministic answer},
    \begin{align}
        &H(A_1^{[k]}|Q_1^{[k]},W_{1:K},\mathcal{R}) = 0 \\
        &H(A_2^{[k]}|Q_2^{[k]},W_{1:K},\mathcal{R}) = 0
    \end{align}
    \item From the first two steps in the CDMS configuration, we obtain the \emph{user privacy} for each database,
    \begin{align}
        I(k;Q_1^{[k]},A_1^{[k]},W_{1:K},\mathcal{R}) = 0 \\
        I(k;Q_2^{[k]},A_2^{[k]},W_{1:K},\mathcal{R}) = 0
    \end{align}
    \item \emph{Validity} becomes \emph{reliability} due to the unique decodable secret $S_k$,
    \begin{align}
        H(W_k|Q_1^{[k]},Q_2^{[k]},A_1^{[k]},A_2^{[k]}) = 0
    \end{align}
    \item \emph{Security} becomes \emph{database privacy} due to the remaining undecodable secrets,
    \begin{align}
        I(W_{\bar{k}};Q_1^{[k]},Q_2^{[k]},A_1^{[k]},A_2^{[k]}) = 0
    \end{align}
\end{enumerate}
\end{Proof}

We are ready to investigate the total communication cost of two-database SPIR by means of the characteristics of CDS and CDMS. We use the terminologies in \cite{Networkscience} for the bipartite graph in CDS/CDMS.

\begin{remark} \label{remark1}
We can construct an upload cost starting from $2 \log_2 K$ in two-database SPIR while satisfying the constraints in the second step of the particular CDMS configuration above. Intuitively, the upload cost of $2 \log_2 K$ comes from the needed $\log_2 K$ bits to be sent to each database to represent any one of the $K$ messages. The upload cost $2 \log_2 K$ can be achieved by the following setting: $X$ and $Y$ are two uniformly selected symbols from a finite set $\mathbb{S}_K = \{0,1,\dots,K-1\}$ such that $X + Y = k - 1$ under an assumption that the sum is always calculated over module $K$. In order to construct a larger upload cost, we can select a larger finite set by utilizing additional dummy messages. As an aside, we note that a larger finite set can be denoted by using multiple symbols from a smaller finite set. This further increases the diversity of upload cost constructions. For example, we can use two symbols from $\mathbb{S}_3 = \{0,1,2\}$ to include every option in $\mathbb{S}_8 = \{0,1,\dots,7\}$. Thus, when $K = 8$, $X$ and $Y$ can either be two one-symbol vectors from $S_8$ or two two-symbol vectors from $S_3$.
\end{remark}

\begin{remark} \label{remark2}
As in CDS and CDMS, we can use a bipartite graph to specify two-database SPIR constraints. As introduced in \cite{CDS, linearCDS}, CDS can be viewed as a data storage system over a bipartite graph where the nodes in each side of the graph are used to denote the input values in each party, and the connectivity of the links is used to indicate the satisfaction of the condition after selecting two nodes (input values) from two parties. In the extension to CDMS, we assign a distinct color $c_k$ to each independent secret $S_k$. Hence, in CDMS, the color of links is used to indicate which secret should be revealed while keeping all the other secrets completely private. Following CDMS, in two-database SPIR, the nodes are used to denote the queries received by the databases, and the links with different colors are used to indicate which message should be retrieved while keeping all the other messages completely private, which implies reliability and database privacy. 
\end{remark}

\begin{remark} \label{remark3}
In the bipartite graph, the links that are incident to any node should include all possible colors with equal number, due to user privacy. 
\end{remark}

\begin{example}
    In this example, we will show the use of bipartite graphs for SPIR for $N = 2$, $K = 3$ and two example upload costs of $U=2 \log_2 3$ and $U=4$. We use colors red, yellow and green to denote messages $W_1, W_2, W_3$, respectively. 
    
    For upload cost of $U=2 \log_2 3$, we use one-symbol vectors $X$ and $Y$ where $X$ and $Y$ are both uniformly selected from $\mathbb{S}_3$ s.t. $X + Y = k - 1$ for message $k$. In this case, globally known condition functions are set accordingly as: $f_i(X,Y) = X + Y + 2 - i$, for $i \in [3]$. Then, we use $A_0, A_1, A_2$ to denote the three choices for the queries in database 1, and $B_0, B_1, B_2$ to denote the three choices for the queries in database 2. The corresponding bipartite graph is shown in Fig.~\ref{K3U2log3_alt}. 
    \begin{figure}[ht]
	    \centering
	    \includegraphics[width=0.5\columnwidth]{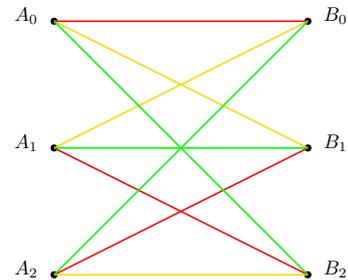}
	    \caption{Bipartite graph for $K = 3$ messages and $U = 2 \log_2 3$ upload cost.}
	    \label{K3U2log3_alt}
    \end{figure}
    
    For upload cost of $U=4$, we use two-symbol vectors $X = \{X_2,X_1\}$ and $Y = \{Y_2, Y_1\}$ where $X_1, X_2, Y_1, Y_2$ are all uniformly selected from $\mathbb{S}_2$ s.t. $2(X_2+Y_2) + (X_1+Y_1) = k - 1$ for message $k$. The setting of globally known condition functions is similar: $f_i(X,Y) = 2(X_2+Y_2) + (X_1+Y_1) + 2 - i$, for $i \in [3]$. Then, we use $A_{00}, A_{01}, A_{10}, A_{11}$ and $B_{00}, B_{01}, B_{10}, B_{11}$ to denote the choices for the queries in the two databases. The corresponding bipartite graph is shown in Fig.~\ref{K3U4_alt}.
    \begin{figure}[ht]
	    \centering
	    \includegraphics[width=0.5\columnwidth]{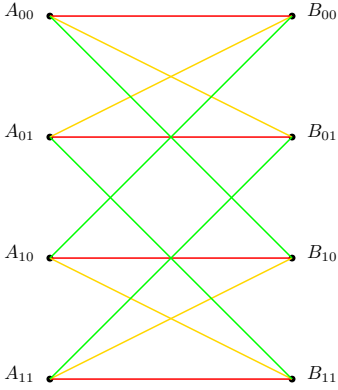}
	    \caption{Bipartite graph for $K = 3$ messages and $U = 4$ upload cost.}
	    \label{K3U4_alt}
    \end{figure}
    \vspace{-1em}
\end{example}

\begin{remark} \label{remark4}
Given an achievable scheme for two-database SPIR with $K = P$ messages with known upload cost $U$ and download cost $D$, we can construct a new achievable scheme for $K = 2P$ messages with upload cost $U + 2$ and download cost $2D$. We use the following simple example to illustrate the idea of the general construction.
\end{remark}

\begin{example}
    Consider two-database SPIR with $K = 4$ messages, where colors red, yellow, green, blue are assigned to messages $W_1, W_2, W_3, W_4$, respectively. Now, first consider a two-database SPIR with $K=2$ messages with a special bipartite graph provided in Fig.~\ref{K2U2_alt}. Following this bipartite graph, we generate an SPIR achievable scheme for $K = 2$ and $L = 1$, with $U = 2$ and $D = 2$ as follows:
    \begin{align}
        &A_0 = S_1,  & B_0 = W_1 + S_1 \\
        &A_1 = W_1+W_2+S_1, & B_1 = W_2+S_1 
    \end{align}
    \vspace{-1.5em}
    
    \begin{figure}[ht]
	    \centering
	    \includegraphics[width=0.5\columnwidth]{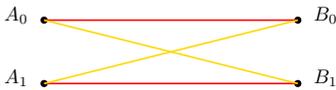}
	    \caption{Bipartite graph for $K = 2$ messages and $U = 2$ upload cost.}
	    \label{K2U2_alt}
    \end{figure}
    
    Now, we use the bipartite graph in Fig.~\ref{K2U2_alt} as a building block to construct an SPIR scheme for $K = 4$ messages as stated in Remark~\ref{remark4}. First, we replicate this bipartite graph, thus, we need to use one extra bit to describe the query choices in each database, see the left part of Fig.~\ref{K4U4_expansion}. Then, we replicate the whole left part, change the color of links to green and blue, and then also exchange the order of query choices in the second column, see the right part of Fig.~\ref{K4U4_expansion}. Combining the left part and the right part in Fig.~\ref{K4U4_expansion}, we can verify that this new bipartite graph is a valid one by checking Remark~\ref{remark2} and Remark~\ref{remark3}. Moreover, following this bipartite graph for $K = 4$, the corresponding upload cost increases by $2$ and the corresponding download cost doubles; see the following achievable scheme with $L = 1$: 
    \begin{align}
        &A_{00} = \{S_1,S_3\}, \quad B_{00} = \{W_1+S_1,W_3+S_4\} \\
        &A_{01} = \{W_1+W_2+S_1,W_3+W_4+S_3\},  \\
        &B_{01} = \{W_2+S_1,W_4+S_4\} \\
        &A_{10} = \{S_2,S_4\}, \quad B_{10} = \{W_1+S_2,W_3+S_3\} \\
        &A_{11} = \{W_1+W_2+S_2,W_3+W_4+S_4\}, \\
        &B_{11} = \{W_2+S_2,W_4+S_3\} 
    \end{align}
    
    \begin{figure}[ht]
	    \centering
	    \includegraphics[width=0.9\columnwidth]{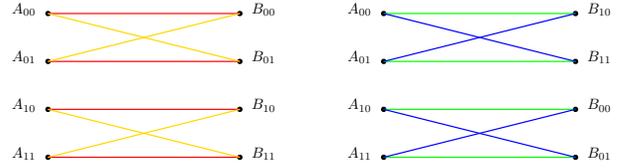}
	    \caption{Bipartite graph for $K = 4$ messages and $U = 4$ upload cost.}
	    \label{K4U4_expansion}
    \end{figure}
\end{example}

\section{Exact Upload-Download Region $N = 2$, $K = 3$}
In this section, we give the exact achievable $(U,D)$ cost region of two-database SPIR for $K=3$ messages using the results of the previous section. In particular, for the upload cost of $U = 2 \log_2 3$, we achieve a download cost of $D = 3$. This outperforms the best-known result of $D = 4$ in \cite{Min_Uploadcost_SPIR}. We show $(2 \log_2 3,3)$ corner point to be optimum with a converse. Further, by increasing the query selection for each database by one, we achieve a download cost of $D = 2$. This means that $U = 4$ is sufficient to achieve $D = 2$, and having $U=6$ is not necessary as in \cite{SPIR}. We show $(4,2)$ corner point to be optimum as well with a converse.

\begin{theorem} \label{thm3}
In two-database SPIR with $K = 3$ messages with message length $L$, when the upload cost is $U = 2 \log_2 3$, the optimal download cost is $D=3L$ and the minimal amount of required common randomness is $2L$.
\end{theorem}

\begin{corollary} 
In two-database SPIR with $K = 3$ messages, if the message length is confined to be $L = 1$, the optimal total communication cost is $2 \log_2 3 + 3$ with minimal amount of required common randomness being $2$.
\end{corollary}

\begin{Proof}
We present the converse proof first. First, we select two random nodes from the two columns. Without loss of generality, let them be $A_1$ and $B_1$, respectively. From $A_1, B_1$, we can recover one random message $W_p$ without learning anything about the remaining messages $W_{\bar{p}}$. Next, we select another two nodes $A_i$, $i \neq 1$ and $B_j$, $j \neq 1$ such that $W_p$ can be recovered from $A_i, B_j$ once again with no knowledge about $W_{\bar{p}}$. Thus, from $A_i, B_1$, we can only recover another random message $W_q, q \neq p$. Then, we have,
\begin{align}
    &H(A_1|\mathcal{F}) + H(B_1|\mathcal{F}) \notag \\ 
    &\geq H(A_1|A_i,B_1,\mathcal{F}) + H(B_1|A_i,B_j,\mathcal{F}) \\
    &= H(A_1,A_i,B_1,\mathcal{F}) + H(A_i,B_1,B_j,\mathcal{F}) \notag \\
    &\quad - H(A_i,B_1,\mathcal{F}) - H(A_i,B_j,\mathcal{F})  \label{proof1.0} \\
    &= H(W_p,A_1,A_i,B_1,\mathcal{F}) + H(W_p,A_i,B_1,B_j,\mathcal{F}) \notag \\
    &\quad - H(A_i,B_1,\mathcal{F}) - H(A_i,B_j,\mathcal{F}) \label{proof1.1} \\
    &\geq H(W_p,A_i,B_1,\mathcal{F}) + H(W_p,A_1,A_i,B_1,B_j,\mathcal{F}) \notag \\
    &\quad - H(A_i,B_1,\mathcal{F}) - H(A_i,B_j,\mathcal{F}) \label{proof1.2} \\
    &= H(W_p) + H(W_p,A_1,A_i,B_1,B_j,\mathcal{F}) - H(A_i,B_j,\mathcal{F}) \label{proof1.3} \\
    &= H(W_p) + H(W_{\bar{p}},A_1,A_i,B_1,B_j,\mathcal{F}) - H(A_i,B_j,\mathcal{F}) \label{proof1.4} \\
    &\geq H(W_p) + H(W_{\bar{p}},A_i,B_j,\mathcal{F}) - H(A_i,B_j,\mathcal{F})  \\
    &= H(W_p) + H(W_{\bar{p}}) \label{proof1.5} \\
    &= 3L
\end{align}
where \eqref{proof1.1} follows from the decodability of message $W_p$ from $A_1, B_1$ and from $A_i, B_j$, \eqref{proof1.2} follows form the fact that conditioning cannot increase entropy, i.e., $H(A_1|W_p,A_i,B_1,\mathcal{F}) \geq H(A_1|W_p,A_i,B_1,B_j,\mathcal{F})$, \eqref{proof1.3} and \eqref{proof1.5} both come from the database privacy \eqref{Database Privacy}, \eqref{proof1.4} follows from the fact that we can always decode $W_{1:3}$ from $A_1, B_1, A_i, B_j, \mathcal{F}$, which can be readily proved by contradiction in the bipartite graph. As a consequence, we reach the desired converse result for the download cost,
\begin{align} \label{conv1}
    D \geq H(A_1) + H(B_1) \geq H(A_1|\mathcal{F}) + H(B_1|\mathcal{F}) \geq 3L
\end{align}

Next, we prove $H(\mathcal{R}) \geq 2L$:
\begin{align}
    0 
    &= I(W_{\bar{p}};A_1,B_1,\mathcal{F}) \\
    &= I(W_{\bar{p}};W_p,A_1,B_1,\mathcal{F}) \label{proof2.1} \\
    &= I(W_{\bar{p}};A_1,B_1|W_p,\mathcal{F}) \label{proof2.2} \\
    &= H(A_1,B_1|W_p,\mathcal{F}) - H(A_1,B_1|W_{1:K},\mathcal{F}) \notag \\
    &\quad + H(A_1,B_1|W_{1:K},\mathcal{F},\mathcal{R}) \label{proof2.3} \\
    &= H(A_1,B_1|W_p,\mathcal{F}) - I(A_1,B_1;\mathcal{R}|W_{1:K},\mathcal{F}) \\
    &\geq H(A_1,B_1|W_p,\mathcal{F}) - H(\mathcal{R}) \label{proof2.4}
\end{align}
where \eqref{proof2.1} follows from the decodability of message $W_p$ from $A_1, B_1$, \eqref{proof2.2} follows from the combination of \eqref{Message IID} and \eqref{Message Set Independence}, \eqref{proof2.3} follows from the deterministic answers \eqref{Deterministic Answer}, and \eqref{proof2.4} follows from \eqref{Message Set Independence} again. Therefore, we turn to find a lower bound for the expression $H(A_1,B_1|W_p,\mathcal{F})$,
\begin{align}
    &H(A_1,B_1|W_p,\mathcal{F}) \notag \\
    &= H(A_1|W_p,B_1,\mathcal{F}) + H(B_1|W_p,\mathcal{F}) \\
    &\geq H(A_1|W_p,A_i,B_1,\mathcal{F}) + H(B_1|W_p,A_i,B_j,\mathcal{F}) \\
    &= H(A_1,A_i,B_1,\mathcal{F}) + H(A_i,B_1,B_j,\mathcal{F}) \\
    &\quad - H(W_p,A_i,B_1,\mathcal{F}) - H(A_i,B_j,\mathcal{F}) \\
    &= H(A_1,A_i,B_1,\mathcal{F}) + H(A_i,B_1,B_j,\mathcal{F}) \\
    &\quad - H(A_i,B_1,\mathcal{F}) - H(A_i,B_j,\mathcal{F}) - H(W_p) \\
    &\geq H(W_p) + H(W_{\bar{p}}) - H(W_p) \label{proof3.1}\\
    &= 2L
\end{align}
where \eqref{proof3.1} exactly follows from the steps between \eqref{proof1.0}-\eqref{proof1.5}. As a consequence, we reach the desired converse result for the minimal amount of required common randomness,
\begin{align} \label{conv2}
   H(\mathcal{R}) \geq 2L
\end{align}

Next, we proceed to the achievability. We use the structure in Fig.~\ref{K3U2log3_alt} and the corresponding answers for $L = 1$ as follows (we use this achievable scheme multiple times for larger $L$),
\begin{align}
    &A_0 \!=\! (S_1, S_2), \; B_0 \!=\! W_1\!+\!S_1 \label{ach1}\\
    &A_1 \!=\! (W_1\!+\!W_2\!+\!S_1, W_2\!+\!W_3\!+\!S_2), \; B_1 \!=\! W_2\!+\!S_2 \label{ach2} \\
    &A_2 \!=\! (W_1\!+\!W_3\!+\!S_1, W_1\!+\!W_2\!+\!S_2), \; B_2 \!=\! W_3\!+\!S_1\!+\!S_2 \!\!\! \label{ach3}
\end{align}
The achievability in (\ref{ach1})-(\ref{ach3}) together with converses in (\ref{conv1}) and (\ref{conv2}) complete the proof.
\end{Proof}

\begin{theorem} \label{thm4}
In the two-database SPIR with $K = 3$ messages with message length $L$, when the upload cost is $U = 2 \log_2 4 = 4$, the optimal download cost is $D=2L$ and the minimal amount of required common randomness is $L$.
\end{theorem}

\begin{corollary} 
In the two-database SPIR with $K=3$ messages, if the message length is confined to be $L = 1$, the optimal total communication cost is $4+2=6$ with minimal amount of required common randomness being $1$.
\end{corollary}

\begin{Proof}
     The converse proof comes from \cite[Thm.~1]{SPIR}. The achievability comes from the following answers with the structure in Fig.~\ref{K3U4_alt},
     \begin{align}
        &A_{00} = S_1, \; B_{00} = W_1 + S_1 \\
        &A_{01} = W_1+W_2+S_1, \; B_{01} = W_2+S_1 \\
        &A_{10} = W_1+W_3+S_1, \; B_{10} = W_3+S_1 \\
        &A_{11} = W_2+W_3+S_1, \; B_{11} = W_1+W_2+W_3+S_1
    \end{align}
    which completes the proof.
\end{Proof}

\begin{figure}[t] 
	\centering
	\includegraphics[width=0.67\columnwidth]{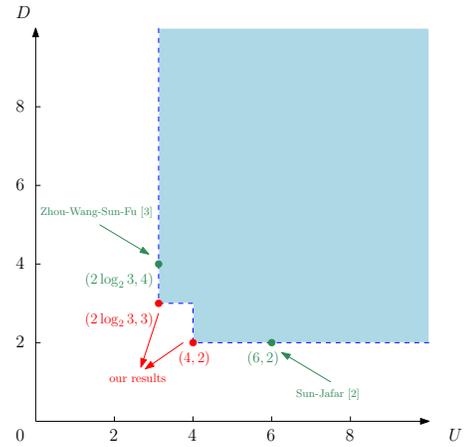}
	\vspace*{-0.2cm}
	\caption{Achievable $(U,D)$ region for two-database SPIR with $K = 3$, $L = 1$.}
	\label{K3L1_Achievable}
	\vspace*{-0.3cm}
\end{figure} 

Combining Theorem~\ref{thm3} and Theorem~\ref{thm4}, we obtain the achievable $(U,D)$ region for two-database SPIR for $K = 3$ and $L = 1$ in Fig.~\ref{K3L1_Achievable}. Any point within the light blue area is achievable, while all the remaining points are not achievable. The optimal communication cost is $4+2=6$. 

\bibliographystyle{unsrt}
\bibliography{ISIT2022}

\end{document}